\documentclass[12pt,preprint]{aastex}

\newcommand{\xmm}{{\it XMM-Newton}}
\newcommand{\puls}{PSR~B2334+61}

\shorttitle{XMM-Newton observations of PSR~B2334+61}
\shortauthors{McGowan et al.}

\received{2005 May 16}
\begin{document}

\title{Evidence for surface cooling emission in the XMM-Newton spectrum 
of the X-ray pulsar PSR~B2334+61}

\author{K.E. McGowan\altaffilmark{1,2}, S. Zane\altaffilmark{1}, M. Cropper\altaffilmark{1}, W.T. Vestrand\altaffilmark{2}, C. Ho\altaffilmark{2}}
\altaffiltext{1} {Mullard Space Science Laboratory, University College of
London, UK}
\altaffiltext{2} {Los Alamos National Laboratory, Los Alamos, NM 87545}
\email{km2@mssl.ucl.ac.uk}

\begin{abstract}

We report on the first \xmm\ observation of the Vela-like pulsar
\puls. Spectral analysis reveals soft X-ray emission, with the
bulk of the photons emitted at energies below $\sim 1.5$~keV.  We
find that the spectrum has a thermal origin and is well-fitted
with either a blackbody or a magnetized, pure H atmospheric model. In 
the latter case, for a neutron star with a radius of $13$~km and a magnetic 
field of $10^{13}$ G, the best-fit gives an hydrogen column density 
$N_{H}=0.33 \times 10^{22}$ cm$^{-2}$ and an effective temperature 
$T_{eff}^\infty= 0.65 \times 10^{6}$~K, as measured at Earth. A comparison 
of the surface temperature of \puls\ obtained from this fit with cooling curves
favor a medium mass neutron star with $M \sim 1.45 M_\sun $ or $ M
\sim 1.6 M_\sun$, depending on two different models of proton
superfluidity in the interior. We do not detect any pulsed emission from the 
source, and determine an upper limit of 5\% for the modulation amplitude of 
the emission on the pulsar's radio frequency.

\end{abstract}


\section{Introduction}
\label{intro}

Young pulsars, neutron stars in supernova remnants (SNRs) and cooling 
neutron stars make up a subset of rotation-powered radio pulsars that can 
be observable at shorter wavelengths.  Measurements of the high energy
radiation are crucial as they provide important insights into the emission
processes and the neutron star physics.

Through observation and theoretical modeling it is generally accepted that a 
combination of emission mechanisms are responsible for the detected X-ray flux 
from rotation-powered pulsars \citep[for a review see e.g.][]{bec01,bec02}.  Soft 
thermal radiation is produced by cooling of the surface of the neutron star \citep[cf.][]{gre83,rom87,pav95}, while the acceleration of particles in the neutron 
star magnetosphere generates non-thermal radiation \citep[cf.][and references 
therein]{mic91,bes93}.  A harder thermal component can be produced by hot 
spots at the polar regions \citep{kun93,zav95}, and the presence of a 
synchrotron nebula may contribute another source of non-thermal radiation 
\citep[cf.][]{aron93}.

The dominant emission mechanism is related to the age of the pulsar \citep[see 
e.g.][]{bec96,pa02}.  In pulsars older than $10^{6}$~yr standard cooling scenarios 
predict that the temperature of the neutron star surface will be too low to generate
detectable thermal X-ray emission \citep{nom87}.  However, the source may be 
observable if reheating via rotational energy loss or heating of the polar caps 
takes place \citep{brin85}.  The middle-aged pulsars ($10^4-10^6$~yr) exhibit 
spectra that can be described by thermal emission from the surface of the 
neutron star.  In some cases a power-law component is also detected 
\citep[e.g.][]{bec02,pa02}.  In pulsars younger than $10^{4}$~yr the strong 
magnetospheric emission prevails over the weaker thermal radiation.    By 
comparing measurements of the soft X-ray flux emitted from the neutron star 
surface with thermal evolution calculations it is possible to investigate the 
physical processes that occur in the interiors of these objects \citep{sch99}.

Although many rotation-powered pulsars have now been observed in X-rays,
when one considers the number of known radio pulsars, the sample of sources 
detected at X-ray energies is still small.  The prospects of detecting a greater 
number of these objects is however improving with missions like \xmm\ and 
{\it Chandra}.  It is now possible, even for faint and distant objects like the one 
presented in this paper, to discriminate whether the dominant mechanism is 
thermal or non-thermal which in turn permits the first estimates of temperature 
and/or power law index.  In particular, the detection of emission due to cooling 
is essential to determine where the pulsar lies on the thermal/evolutionary 
diagram, with the main objective being able to constrain observationally the 
equation of state of matter at supra-nuclear densities.  Currently, there are very 
few sources for which this is possible, therefore any new source that can be 
added to the parameter space is important.

One such object that should fall into the category of pulsars for which the cooling
neutron star is the dominant source of emission is \puls.  This Vela-like pulsar is
located at a distance of $D=3.1^{+0.2}_{-1.0}$ kpc \citep{cor02}.  Analysis 
by \citet{kul93} implies that the pulsar is associated with the SNR G114.3, making 
the source one of the oldest (spin-down age of $\sim 4\times 10^{4}$~yr) that is 
still linked to a SNR.  Hence, \puls\ is a key object to study with relation to the
thermal/evolutionary parameter space.  The spin-down rate 
($\dot P = 191\times 10^{-15}$ s s$^{-1}$) of the 495 ms pulsar indicates that 
the magnetic field of \puls\ is $B\sim 10^{13}$ G.  \puls\ was originally detected in 
a short (8 ks) {\it ROSAT} pointing \citep{bec93,bec96} but the low statistics 
prevented a comprehensive investigation of the spectral properties of the source 
or a meaningful temporal analysis.  Here we report on the first \xmm\ observation 
of \puls.

\section{Observations and data reduction}
\label{obser}

\puls\ was observed with \xmm\ on 2004 February 12.  For the spectral and
timing analysis we used data from the European Photon Imaging Camera (EPIC)
instruments: the EPIC MOS detector \citep{tur01} and the EPIC-PN detector
\citep{str01}.  Both MOS instruments and the PN were configured in {\it
full frame} mode and we used the thin filter.  The MOS1 and MOS2 observations 
had total exposure times of 47.5 ks.  The EPIC-PN buffer was filled during the 
observation resulting in an exposure time of 45.8 ks.  We reduced the EPIC
data with the \xmm\ Science Analysis System (SAS version 6.0.0).

In order to maximize the signal-to-noise ratio for our \xmm\ observation,
we filtered the data to include only single, double, triple and quadruple
photon events for the MOS, and only single and double photon events for
the PN.  We included photons with energies only in the range $0.3-10$ keV.

To verify that there is no contribution to the measured emission from
the remnant, we compared the radial profile of the pulsar emission with the 
\xmm\ point-spread function for EPIC-PN at 1.5 keV, which we generated using 
the King profile parameters included in the \xmm\ calibration file 
''XRT3\_XPSF\_0006.CCF.plt''
\footnote{See http://xmm.vilspa.esa.es/docs/documents/CAL-SRN-0100-0-0.ps.gz 
for more information}.  We find that the emission we detect from \puls\ is 
consistent with that from a point source.

Data have been extracted using a circular region of radius $30\arcsec$,  
centered on the radio pulsar's position.  We obtained 600 counts from MOS1, 
616 from MOS2, and 2384 from PN.  By using a region of equal size offset from 
the pulsar's position, we found that the background contributes 439 counts in 
MOS1, 449 in MOS2, and 1846 in PN.  We detect X-rays from \puls\ at a level of
$5\sigma$ in MOS1, $5.1\sigma$ in MOS2, and $8.3\sigma$ in PN.  The extraction 
radius we used encircles 85\% of the energy from the source in MOS and 83\% in 
PN.  The measured fluxes reported in Table \ref{tab_spec} have been corrected 
accordingly.

\section{Spectral Analysis}
\label{spec}

In order to perform the spectral analysis, we extracted the pulsar spectrum 
from the MOS1, MOS2 and PN event files.  Data were filtered to exclude events 
that may be incorrect, for example those next to the edges of the CCDs and 
next to bad pixels.  The three spectra were regrouped by requiring at least 
50 counts per spectral bin.  For each spectrum we created a photon 
redistribution matrix (RMF) and ancillary region file (ARF). The subsequent 
spectral fitting and analysis was performed using XSPEC, version 11.3.1.

\begin{figure}
\epsscale{0.6}
\plotone{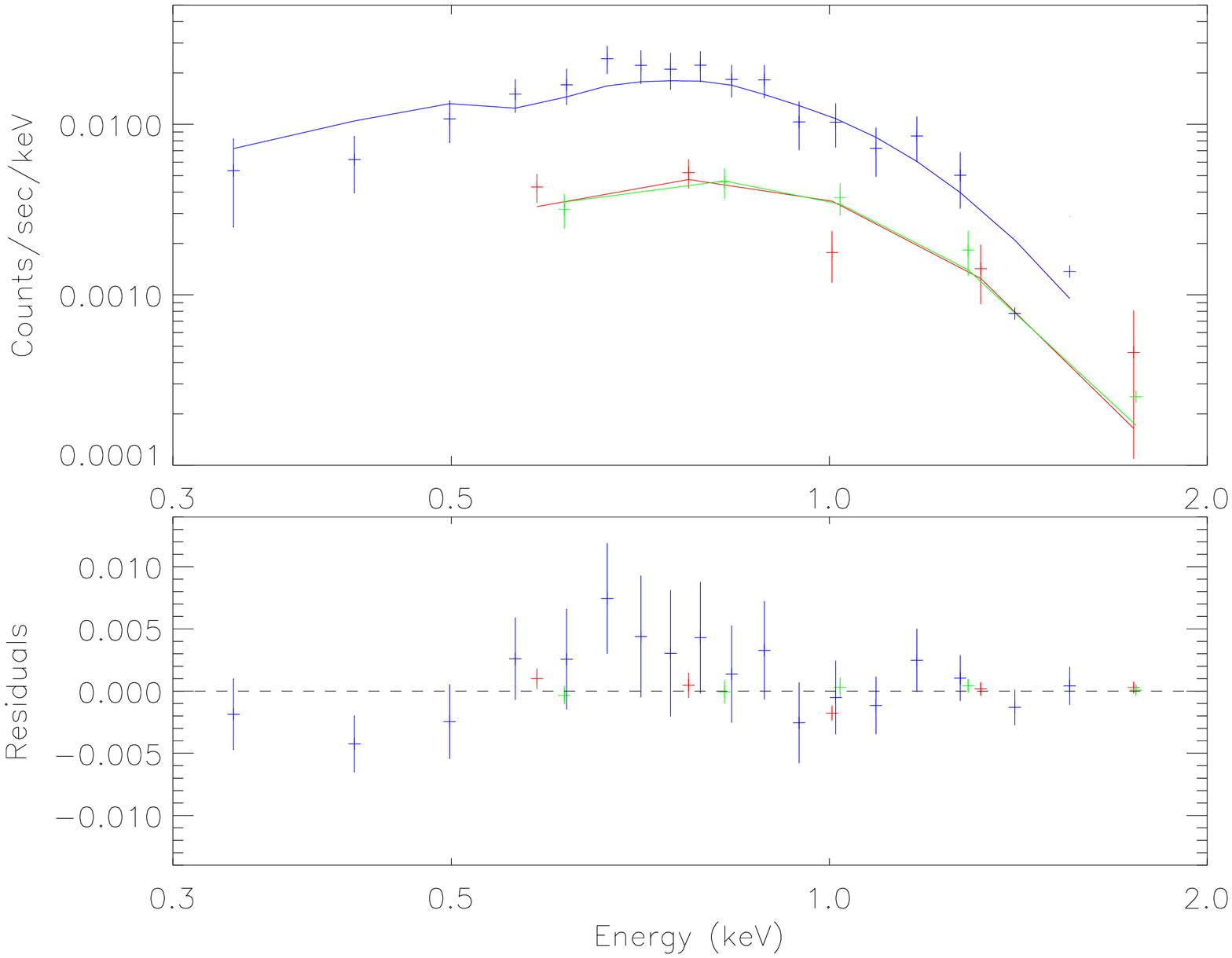}
\plotone{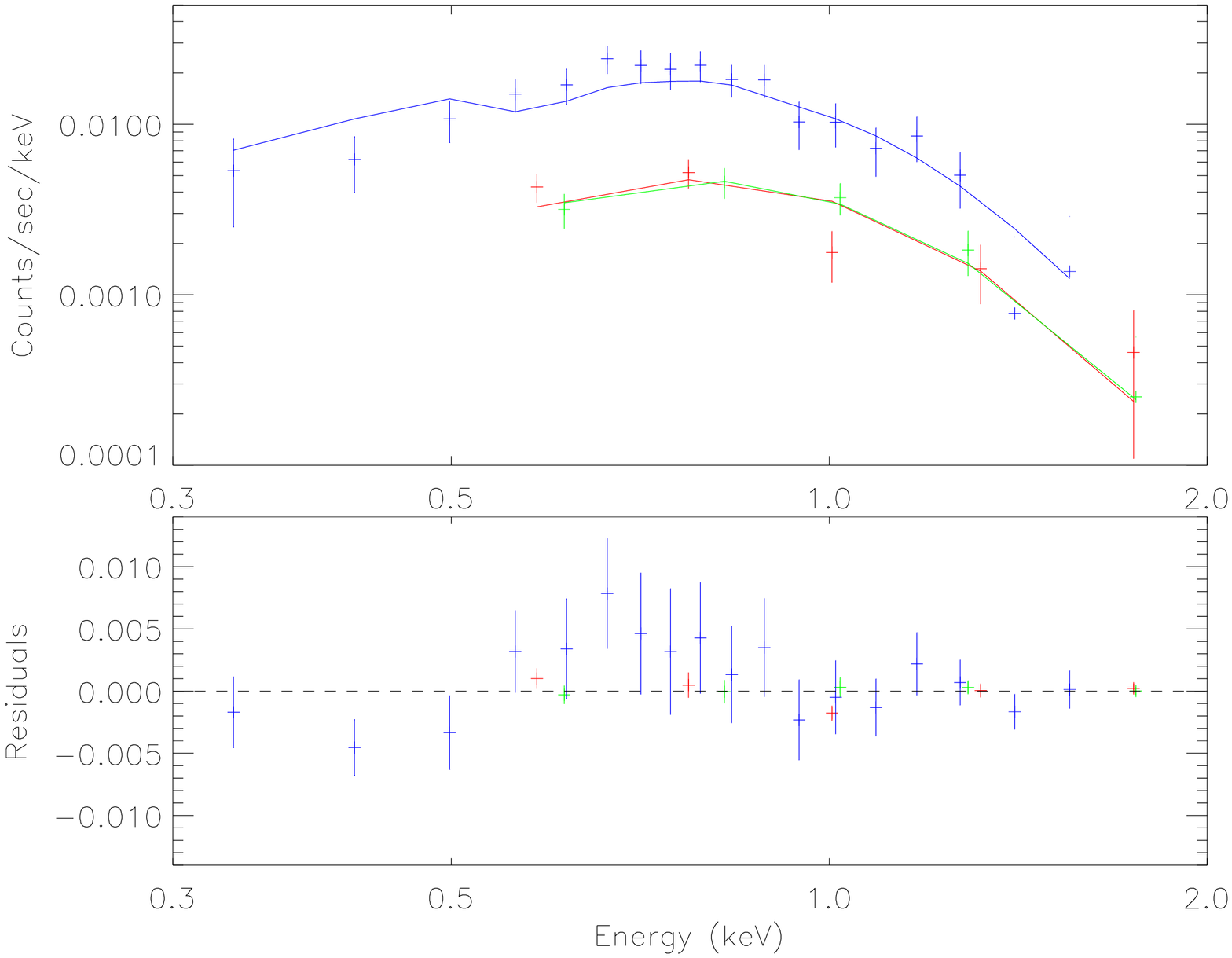}
\caption{\xmm\ PN and MOS spectra of the X-ray emission detected
from \puls. First panel: data (crosses) and best-fit blackbody
model (solid line) for the parameters given in
Table~\ref{tab_spec}. Second panel: Difference between the data
and the blackbody model. Third panel: data (crosses) and best-fit
magnetized, pure H, atmospheric model (solid line) for the
parameters given in Table~\ref{tab_spec} (NSA$^{\rm b}$). Fourth
panel: Difference between the data and the atmospheric model.  In the 
electronic edition the PN, MOS1 and MOS2 data points, and best fit model 
where appropriate, are colored blue, red and green, respectively. }
\label{fig_spec}
\end{figure}

Since essentially no emission is detected at energies  $>1.5$ keV, we modeled 
the MOS spectra over $0.5-1.5$ keV and the PN spectrum over $0.3-1.5$ keV 
(taking into account the calibration uncertainties for the MOS at low 
energies).  We fitted the combined MOS and PN spectra using different models: 
initially a power-law, a blackbody and a combination of both, each modified 
by photoelectric absorption (wabs in XSPEC). The best-fit results are 
summarized in Table \ref{tab_spec}.  The $\chi^{2}$ values indicate that the 
spectrum is better represented  by a thermal (blackbody) model, rather than by 
an absorbed power-law.  The additional power law component in the third (and 
fifth) fit is poorly constrained, and not required: by applying an $F$-test, we
find an $F$-statistic of 1.6 and probability of 0.2.  The best-fit blackbody 
fit gives $N_{H} = 0.26 \times 10^{22}$ cm$^{-2}$ and 
$T^\infty = 1.62 \times 10^6$ K (where $T^\infty$ is the blackbody 
temperature measured ad Earth\footnote{We note that $T_{eff}^\infty$ can be 
obtained by red-shifting at Earth the effective temperature measured at the 
neutron star surface, $T_{eff}$, through the relation 
$T_{eff}^\infty = T_{eff}(1-2.952M_{ns}/R)^{0.5}$.  $T_{eff}$ is one of the 
best-fitting  parameters of the NSA model.}; see Figure~\ref{fig_spec}, first 
and second panels).  We notice that, consistently, this value for the hydrogen 
column density is lower than the Galactic one in the source direction, which 
is $N_H \sim  0.84 \times 10^{22}$  cm$^{-2}$, while the same argument can 
only be applied marginally (within the lower limit of the error bars) to the 
value of $N_H$ inferred from the power-law fit.  The radius of the emitting 
region, as derived from the blackbody fit using the \citet{cor02} distance, 
is $R_{em}=1.66^{+0.59}_{-0.39}$~km.

In order to attempt a different representation of the thermal emission 
detected from \puls, we used a magnetized, pure H atmospheric model (nsa in 
XSPEC, for details see \citealt{pav95}), by fixing the magnetic field at 
$B=10^{13}$~G, the value inferred from the radio timing measurements of the 
source.  The neutron star mass and radius (measured at the source) were fixed 
at $M_{ns}= 1.4 M\sun$ and $R=10$~km, respectively, although the spectral fit 
is not particularly sensitive to these parameters (see \S\ref{disc} for an a 
posteriori estimate of the mass).  We found that this model also provides an 
excellent representation of the pulsar spectrum, and the best-fit parameters are 
$N_{H} = 0.42 \times 10^{22}$ cm$^{-2}$ and $T_{eff}^\infty = 0.58 \times 
10^{6}$ K, where $T_{eff}^\infty$ is the effective temperature as measured at 
Earth$^2$ (see Table \ref{tab_spec}).  However, this fit gives a pulsar distance 
of $D=1.1\pm0.6$ kpc, lower than that inferred from the value of the electron 
density ($D=3.1^{+0.2}_{-1.0}$ kpc, \citealt{cor02}). Although the pulsar distance 
and radius are poorly constrained by our data, we find that a better agreement 
can be reached by using the same atmospheric model and fixing the neutron 
star radius at a slightly larger value of $13$ km.  In this case, we find 
$N_{H}= 0.33\times 10^{22}$ cm$^{-2}$ and $T_{eff}^\infty= 0.65\times 
10^{6}$~K (see Table \ref{tab_spec} and Figure~\ref{fig_spec}, third and fourth 
panels).  The distance resulting from this fit is $D=3.2\pm 1.7$ kpc, which is 
consistent with the electron density distance. Finally, we verified that even 
when the thermal component is modeled by a neutron star atmosphere, an 
additional power law is not required (last fit in Table \ref{tab_spec}).

We note that the residuals from the model fits shown in Figure~\ref{fig_spec} 
are relatively large at $\sim 0.6-0.7$ keV.  To try and reduce the residuals 
we fitted the data using the previous models, each modified by an absorption, 
namely phabs or tbabs in XSPEC.  We find that the resulting fits are not 
improved with respect to our previous results.

\section{Timing Analysis}
\label{timing}

For the timing analysis we only used PN data, extracted by applying the
filtering criteria and extraction region previously described in 
\S\ref{obser}; the filtered file was then barycentrically corrected.  In 
order to search for an X-ray modulation at the \puls\ spin period, we first 
determined a predicted pulse period at the epoch of our \xmm\ observations, 
assuming a linear spin-down rate and using the radio measurements 
\citep{dtws85,hlk04}. We calculate $P = 0.495355469$ s ($f = 2.0187523$ Hz) 
at the midpoint of our observation (MJD 53,047.6).  As glitches and/or 
deviations from a linear spin-down may alter the period evolution, we then 
searched for a pulsed signal over a wider frequency range centered on 
$f = 2.01875$ Hz.  We searched for pulsed emission using two methods.  In the 
first method we implement the $Z^{2}_{n}$ test \citep{buc83}, with the number 
of harmonics $n$ being varied from 1 to 5. In the second method we calculate 
the Rayleigh statistic \citep{dej91,mar72} and then calculate the maximum 
likelihood periodogram (MLP, \citealt{za02}) using the $C$ statistic 
\citep{cas79} to determine significant periodicities in the data sets.

We do not find any significant peak near to the predicted frequency with 
either method, either using the whole $0.3-10$~keV energy band or restricting 
the search to the $0.3-1.5$~keV band.  By folding the light curve of \puls\ 
on the radio frequency and fitting it with a sinusoid, we determine an upper 
limit for the pulsation of 5\% in modulation amplitude (defined as 
$(F_{max}-F_{min})/(F_{max}+F_{min})$ where $F_{max}$ and $F_{min}$ are the 
maximum and minimum of the pulse light curve).

\section{Discussion}
\label{disc}

We have presented the results from the first \xmm\ observation of \puls.  The 
source has been positively detected in all EPIC instruments, although the 
X-ray emission is very faint and the spectrum does not have a high enough
signal-to-noise to make a detailed multi-component fit.  However, single 
component fits allowed us to discriminate between a thermal (blackbody-like) 
or non-thermal (power law-like) nature of the dominant emission mechanism.  
We find that the spectrum is well represented by either thermal 
blackbody-like emission from a small emitting area (e.g. a hot polar cap) of 
size $\sim 2$~km or emission from a pure H atmosphere of a neutron star with 
$R\sim 13$~km and magnetic field $B\sim10^{13}$~G. The EPIC flux, measured 
in the $0.3-10$~keV band, is $(9.2_{-0.9}^{+0.6}) \times 10^{-14}$ erg 
cm$^{-2}$ s$^{-1}$ and $(1.7_{-1.6}^{+0.1}) \times 10^{-13}$ erg cm$^{-2}$ 
s$^{-1}$, in the two cases respectively.  Both values are slightly higher 
than that previously inferred by \cite{bec96}, $(7.1 \pm 0.2) \times 10^{-14}$ 
erg cm$^{-2}$ s$^{-1}$, although we note that due to their limited counts 
{\it ROSAT} data could only be analyzed by making an assumption a priori on 
the spectral shape.  In both cases, the value of $N_H$ is consistently lower 
than the total galactic one in the pulsar direction.  A fit with a power-law 
model alone is statistically worse, and we find little evidence for the
presence of a non-thermal component in the spectrum in addition to the 
thermal one.  The upper limit for the flux contribution from the power-law in 
the energy range $0.3-1.5$ keV to the blackbody fit is 3\%, and to the 
magnetized atmospheric model fit is $<< 1$\%.  We do not detect X-ray 
pulsations corresponding to the radio signal, to a limit of 5\% in modulation 
amplitude.

Although both thermal fits are conceivable options, we tend to prefer the 
atmospheric model representation based on physical grounds.  In this case, 
magnetic effects for a field strength of order $\sim 10^{13}$~G (which is the 
value inferred from measurements of the source spin period and period 
derivative in the radio band) are consistently accounted for in the radiative 
transfer computation.  Moreover, a value for the radius of $\sim 10-13$~km is 
in agreement with the prediction of several neutron star equations of state 
\citep{lapra} and we found that this parameter can be successfully adjusted 
to make the distance inferred from the spectral fit consistent with that 
obtained from the electron density value of this pulsar. 

To date, thermal emission has been detected in only very few radio pulsars: 
PSR~B0656+14 (\citealt{po96}), PSR~B1055-52 (\citealt{pa02}), PSR~J0437-4715 
(\citealt{zavlin02}), PSR~J0538+2817 (\citealt{mcg03}), Geminga 
(\citealt{hw97}), Vela (\citealt{pa01}), and PSR~B1706-44 (\citealt{gott02}, 
\citealt{mcg04}).  In the case of the first three objects the thermal emission
detected above $\sim 0.5$~keV is more likely to originate from a hot-polar 
cap.  These sources are in fact old pulsars, at a more advanced stage of 
their cooling history and their surface emission should peak at much lower 
(UV) energies.  This idea was strengthened by the detection of a further 
thermal component below $0.7$~keV (\citealt{pa02}) in the spectrum of the 
brightest of them, PSR~B0656+14.  The Vela pulsar and PSR~B1706-44 are 
younger ($\tau \sim 10^4$~yrs) and are the only radio active sources for 
which the thermal component observed in the soft X-rays is well explained by 
a magnetized cooling atmosphere (\citealt{pa01}, \citealt{mcg04}).  When this 
model is assumed instead of a blackbody, the inferred radius is in agreement 
with a neutron star equation of state.  The only other neutron stars whose 
thermal component is better described by an atmospheric model, and for which 
this interpretation resolves all the inconsistencies which follow from the 
blackbody interpretation, are the radio-silent neutron stars 1E~1207-52 
(\citealt{za98}) and RX~J0822-4300 (\citealt{za99}). 

The \xmm\ observation reported here allows us to add another entry to the 
list.  If indeed the X-ray emission detected from \puls\ is originating in 
the cooling atmosphere of the neutron star, our estimate of the effective 
temperature allows us to localize the object in the neutron stars thermal 
evolutionary diagram.

Our knowledge of neutron star interiors is still uncertain and accurate
measurements of the neutron star surface temperature are particularly
important to constrain the cooling models and provide information on the
physics of the neutron star.  Roughly speaking, theoretical models predict
a two-fold behavior of the cooling curves depending on the star mass.  In 
low-mass neutron stars neutrino emission is mainly due to a modified Urca 
process and nucleon-nucleon bremsstrahlung.  These are relatively weak 
mechanisms and produce {\it slow cooling}.  In stars of higher mass the 
neutrino emission is enhanced by a direct Urca process (or other mechanisms 
in exotic matter), therefore these stars cool down much faster ({\it fast 
cooling} regime).  To date (see \citealt{ya02} for a discussion) it has
been realized that simple models which do not account for proton and
neutron superfluidity fail in explaining the surface temperatures observed
in many sources, unless objects such as e.g. Vela, Geminga, RX~J1856-3754
do have exactly the critical mass that bounds the transition between the very 
different {\it slow cooling} and {\it fast cooling} regimes.  This unlikely 
fine-tuning is not required if the effects of nucleon superfluidity are 
accounted for.  In particular, models with proton superfluidity included 
predict an intermediate region between fast cooling and slow cooling curves, 
which is expected to be populated by medium mass neutron stars (roughly with 
$M$ between 1.4 and 1.65 $M_\sun$).  Although the full picture only holds if,
at the same time, neutron superfluidity is assumed to be rather weak, it is 
still interesting that many neutron stars (as 1E~1207-52, Vela, RX~J1856-3754, 
PSR~0656+14) have a surface temperature which falls in such a transition 
region \citep{ya02}.  In turn, this means that measuring the surface 
temperature allows us to ``weigh'' neutron stars \citep{ka01}.  As we can see 
from the first two panels of Figure~2 in \cite{ya02}, assuming an age of
$\log \tau = 4.6$, the surface temperature of \puls\ derived from the 
blackbody fit is even higher than the upper cooling curves i.e. those 
corresponding to the slow cooling regime.  However, the surface temperature 
$\log T^\infty = 5.9$ obtained by fitting with the magnetized model and 
$R=13$~km falls well within the above mentioned transition region of medium 
mass neutron stars.  The mass of \puls\ should then be $\sim 1.45 M_\sun $ 
or $\sim 1.6 M_\sun$, depending on the kind of proton superfluidity assumed 
in the model (1p and 2p respectively).  Indeed, the measured properties of 
\puls\ reported here are in remarkable agreement with that measured for 
PSR~B1706-44 \citep{mcg04}. 

As mentioned in the introduction, we are now in the position to make 
the first experimental studies of surface temperatures of isolated neutron 
stars.  However, although most of our insight of neutron star temperatures 
and interior rely on them, at present these studies have to be considered 
as pioneering.  Only a few sources are currently available for this kind of 
exercise, so every newly discovered candidate is important.  At present, 
X-ray spectra of thermal emission from NSs are fitted with either a blackbody 
or magnetic atmosphere models.  Although the latter definitely represent a 
substantial improvement, inasmuch they include most of the relevant physics, 
a large amount of work remains to be done before they can be claimed to be 
fully satisfactory.  The same is true for the cooling curves: progress has 
been made in improving these models but they can not yet be considered as 
completely realistic.  Besides other effects, we make the caveat that both
spectra and cooling curves are computed assuming 1-dimensional transfer 
of radiation/heat in a single effective temperature, a single magnetic field 
zone and neglect all effects of more realistic neutron star thermal and 
magnetic surface distributions (see \citealt{zt05}, \citealt{pa04}, 
\citealt{bl04}).  However, what values to include for these parameters is not 
necessarily obvious as it is not completely clear what the NS surface 
temperature distribution is even in the case of a simple dipolar magnetic 
field (see \citealt{gkp04}).  We also note that interpreting the temperatures 
obtained from the spectral fits in the context of theoretical cooling curves 
relies on the true age of the pulsar being the same as the characteristic 
spin-down age, which may not be valid.  Our results for \puls\ provide 
information that helps to constrain the current models and will enable more 
realistic models to be produced in the future.

\acknowledgments

This work is based on observations obtained with \xmm, an ESA science mission 
with instruments and contributions directly funded by ESA Member States and 
NASA.  KEM and SZ acknowledge the support of the UK Particle Physics and 
Astronomy Research Council (PPARC).  Part of this work was supported by 
NASA grant R-2865-04-0.  We thank the anonymous referee for a careful reading 
of the manuscript and several helpful comments.  We thank Werner Becker and 
Bernd Aschenbach for their contributions.

\clearpage

\clearpage

\begin{deluxetable}{lccccc}
\tablewidth{0pt}
\tablecaption{Spectral fits to the X-ray emission from PSR B2334+61 
\label{tab_spec}} \tablehead{
\colhead{Model} & \colhead{$N_{H}$}  & \colhead{$\Gamma$} &
\colhead{$T/T_{eff}^\infty$} & \colhead{$\chi^{2}_{\nu}/\nu$}
&\colhead{$F_X$}\\
\colhead{} & \colhead{($10^{22}$ cm$^{-2}$)}  & \colhead{} & \colhead{($10^{6}$ K)}
& \colhead{} & \colhead{(erg cm$^{-2}$s$^{-1}$)} }
\startdata
PL & $0.90^{+0.09}_{-0.24}$ & $8.5^{+1.5}_{-1.8}$ & \nodata & $38/25 = 1.5$ & $(6.4_{-5.4}^{+26.1}) \times 10^{-11}$\\
BB & $0.26^{+0.26}_{-0.05}$ & \nodata & $1.62^{+0.23}_{-0.23}$ & $27/25 =
1.1$ & $(9.2_{-0.9}^{+0.6}) \times 10^{-14}$\\
BB+PL & $0.43^{+0.61}_{-0.17}$ & $2.2^{+3.0}_{-1.4}$ &
$1.27^{+0.35}_{-0.58}$ & $24/23 = 1.0$ & $(3.1_{-1.2}^{+1.3}) \times 10^{-13}$\\
NSA$^{\rm a}$ & $0.42^{+0.34}_{-0.04}$ & \nodata &
$0.58^{+0.13}_{-0.25}$ & $28/25 = 1.1$ & $(3.4_{-3.3}^{+6.0}) \times 10^{-13}$\\
NSA$^{\rm b}$ & $0.33^{+0.41}_{-0.10}$ & \nodata & $0.65^{+0.13}_{-0.34}$
& $31/25 = 1.2$ & $(1.7_{-1.6}^{+0.1}) \times 10^{-13}$\\
NSA$^{\rm b}$+PL & $0.26^{+0.73}_{-0.04}$ &
$9.4^{+0.6}_{-12.0}$ &
$0.76^{+0.09}_{-0.65}$ & $34/23 = 1.5$ & $(1.0_{-1.0}^{+0.7}) \times 10^{-13}$ \\
\enddata
\tablecomments{Uncertainties are all at 90\% confidence. Models
used: ``PL'' indicates an absorbed power law with photon index
$\Gamma$; ``BB'' indicates an absorbed blackbody emitting at a
temperature $T$; ``NSA'' indicates an absorbed, magnetized, pure H
atmospheric model with $B= 10^{13}$ G and effective temperature 
$T_{eff}^\infty$, as measured at Earth \citep{pav95}. For the NSA 
model we fixed the neutron star mass at $M_{\rm NS}= 1.4 M_{\sun}$, 
and the radius at 10~km (NSA$^{\rm a}$) or 13~km (NSA$^{\rm b}$). 
The last column is the unabsorbed flux in the $0.3-10$~keV band.
\\}
\end{deluxetable}

\end{document}